\documentclass[sort&compress,final,3p,times,twocolumn,fixfloat]{elsarticle}

\usepackage{graphicx}
\usepackage{amssymb} 
\usepackage{amsmath} 
\usepackage{times}
\usepackage{tabularx}

\usepackage{txfonts}
\usepackage{lineno}

\journal{Physics Letters B}

 \usepackage{ifpdf}
 \ifpdf
 \usepackage[pdftex]{hyperref}
 \else
 \usepackage[hypertex]{hyperref}
 \fi

 \hypersetup{
   pdftitle={},%
   pdfauthor={},%
   pdfsubject={},%
   pdfkeywords={},%
   pdfstartview={},%
   bookmarksopen=true, breaklinks=true, debug=true, %
   colorlinks=true, linkcolor=red, citecolor=blue, urlcolor=blue
 }

\usepackage{float} 
\usepackage[caption=false]{subfig} 

\newcount\savefnused
\newcount\savefndone

\newcommand{\savefootnote}[2][\empty]
{\ifx\empty#1\footnotemark\else\footnotemark[#1]\fi
 \global\advance\savefnused by 1
 \expandafter\xdef\csname savefnmark\the\savefnused\endcsname{\thefootnote}%
 \expandafter\xdef\csname savefntext\the\savefnused\endcsname{#2}%
}
\newcommand{\flushfootnote}{\loop\ifnum\savefndone<\savefnused
  \global\advance\savefndone by 1
  \footnotetext[\csname savefnmark\the\savefndone\endcsname]%
    {\csname savefntext\the\savefndone\endcsname}%
  \global\expandafter\let\csname savefnmark\the\savefndone\endcsname\relax
  \global\expandafter\let\csname savefntext\the\savefndone\endcsname\relax
\repeat}

\usepackage{multirow,bigdelim}

\usepackage{textcomp}
\usepackage{comment}

\usepackage{lineno}

\usepackage{float} 
\usepackage{subfig} 
\usepackage{amsmath} %
\usepackage{amstext}

\usepackage{graphicx}%
\usepackage{tabularx,ragged2e}%

\newcolumntype{Y}{>{\centering\arraybackslash}X}

\newcommand{\eqs}[1]{\begin{equation} \begin{split} #1\end{split} \end{equation} }

\def\ie{{\it i.e.}}
\def\eg{{\it e.g.}}

\def\GeV{{\rm GeV}}
\def\GeV2{{\rm GeV}^2}

\renewcommand{\P}{{\cal P}}

\newcommand{\ce}[1]{Eq.~(\ref{#1})}
\newcommand{\cf}[1]{{Fig.~\ref{#1}}}
\newcommand{\ct}[1]{{Table~\ref{#1}}}

\begin{document} 
\begin{frontmatter}

\title{Indication for Double Parton Scatterings in $W+$Prompt $J/\psi$ Production at the LHC}
\author[a]{Jean-Philippe Lansberg}
\author[b,c]{Hua-Sheng Shao}
\author[a,d]{Nodoka Yamanaka}
\address[a]{IPNO, CNRS-IN2P3, Univ. Paris-Sud, Universit\'e Paris-Saclay, 
91406 Orsay Cedex, France}
\address[b]{Sorbonne Universit\'es, UPMC Univ. Paris 06, UMR 7589, LPTHE, F-75005 Paris, France}
\address[c]{CNRS, UMR 7589, LPTHE, F-75005 Paris, France}
\address[d]{iTHES Research Group, RIKEN, Wako, Saitama 351-0198, Japan}

\begin{abstract}
{\small
We re-analyse the associated production of a prompt $J/\psi$ and a $W$ boson in
 $pp$ collisions at the LHC following the results of the ATLAS Collaboration. We
 perform the first study of the Single-Parton-Scattering (SPS) contributions
at the Next-to-Leading Order (NLO) in $\alpha_s$ in the Colour-Evaporation Model
 (CEM), an approach based on the quark-hadron-duality. Our study provides clear indications
 for Double-Parton-Scattering (DPS) contributions, in particular at low transverse
 momenta, since our SPS CEM evaluation, which can be viewed as a conservative upper limit of
the SPS yields, falls short compared to the ATLAS experimental data by 3.1 standard deviations.
We also determine a finite allowed region for $\sigma_{\rm eff}$, inversely 
proportional to the size of the DPS yields, corresponding to the otherwise opposed
hypotheses, namely our NLO CEM evaluation and the LO direct Colour-Singlet (CS) Model contribution.
In both cases, the resulting DPS yields are significantly larger than that initially assumed 
by ATLAS based on jet-related analyses but is consistent with their observed
raw-yield azimuthal distribution {and with their prompt $J/\psi+J/\psi$ and $Z$ + prompt $J/\psi$
data}.}
\end{abstract}

%
\end{frontmatter}
%

\section{Introduction.}

The simultaneous production of vector bosons and quarkonia at high-energy 
colliders is a very useful observable to study perturbative and nonperturbative 
aspects of Quantum Chromodynamics (QCD). It also provides original means to 
search for new physics beyond the standard model via the Higgs sector, 
as illustrated by the pioneering study of CDF Collaboration~\cite{Acosta:2003mu,Aaltonen:2014rda}.
Recently, the final states $J/\psi+W$~\cite{Aad:2014rua} and 
$J/\psi+Z$~\cite{Aad:2014kba} were observed by the ATLAS Collaboration.
Similar processes with bottomonia have however not been observed 
yet\footnote{Other reactions of specific interest include the production of a 
quarkonium + a photon. It was proposed to constrain the quarkonium-production 
mechanisms~\cite{Roy:1994vb,Mathews:1999ye,Li:2008ym,Lansberg:2009db,Li:2014ava}, 
to study the proton gluon content~\cite{Doncheski:1993dj,Dunnen:2014eta} or to 
probe the $H^0$ coupling to the heavy-quarks~\cite{Doroshenko:1987nj,Bodwin:2013gca}.} .

Besides the associated production with a vector boson, quarkonium-pair production
has been the object of a number of recent experimental studies at the LHC and the 
Tevatron~\cite{Aaij:2011yc,Abazov:2014qba, Khachatryan:2014iia,Aaboud:2016fzt,Aaij:2016bqq}
 --30 years after the pioneering analyses of NA3~\cite{Badier:1982ae,Badier:1985ri}.
At small rapidity separations ($\Delta y_{\psi\psi}$), none of them exhibit any tension with 
the SPS CS contributions (\ie\ the LO in $v^2$ of Non Relativistic QCD 
(NRQCD)~\cite{Bodwin:1994jh}, known up to NLO accuracy~\cite{Lansberg:2013qka,Sun:2014gca,Likhoded:2016zmk})  
whereas they point at a significant DPS contributions for increasing $\Delta y_{\psi\psi}$-- 
in accordance with previously studied observables~\cite{Akesson:1986iv,Alitti:1991rd,Abe:1993rv,Abe:1997xk,Abazov:2009gc,Aad:2013bjm,Chatrchyan:2013xxa}.
Di-$J/\psi$ hadroproduction has in fact been the object of a large number of theoretical works~\cite{Kartvelishvili:1984ur,Humpert:1983yj,Vogt:1995tf,Li:2009ug,Qiao:2009kg,Ko:2010xy,Berezhnoy:2011xy,Li:2013csa,Lansberg:2013qka,Lansberg:2014swa,Sun:2014gca,He:2015qya,Baranov:2015cle,Lansberg:2015lva,Likhoded:2016zmk,Borschensky:2016nkv,Lansberg:2017dzg}. 
The case of $\Upsilon+J/\psi$ production, measured by the D0
 collaboration~\cite{Abazov:2015fbl}, is slightly different as it seems highly
 dominated by DPS contributions. We refer to~\cite{Shao:2016wor} for a complete
 and up-to-date theory discussion of this reaction. Similar conclusions, pointing
 at a dominant DPS yield, were made by LHCb for $J/\psi+$charm~\cite{Aaij:2012dz}
 and $\Upsilon+$charm~\cite{Aaij:2015wpa} production with measured cross sections
significantly larger than the SPS expectations~\cite{Artoisenet:2007xi,Baranov:2006dh,Berezhnoy:2015jga}. We however note that it is
 not clear whether the magnitude of all these DPS yields fit in a coherent picture
 with a {\it universal} $\sigma_{\rm eff}$ which would be inversely proportional to the
 probability of a second parton scattering.

In the case of $J/\psi+Z$ production, the yield observed by  ATLAS happens to be
up to nearly one order of magnitude larger than the theory evaluations 
from NRQCD (with CS and/or Colour-Octet (CO) --higher order in $v^2$--
 contributions)~\cite{Mao:2011kf,Gong:2012ah}.
A natural explanation for such a gap would be --like for quarkonium pairs at large
 $\Delta y_{\psi\psi}$-- associated production through DPS  but for the fact that the
 azimuthal distribution of the observed events shows a significant back-to-back
 peak pointing at a large SPS. In a recent paper~\cite{Lansberg:2016rcx}, we
 have solved this apparent conflict by invoking a misinterpretation
of this raw-yield azimuthal distribution {when it results from} two different sources with very different
transverse-momentum distributions in a detector like ATLAS with a strongly
transverse-momentum-dependent acceptance. In a further study~\cite{Lansberg:2016muq},
 we have shown that the production of a $Z$ with a non-prompt $J/\psi$ (\ie\ from a
 $b$-hadron decay), as reported by ATLAS in~\cite{Aad:2014kba}, is well accounted by
the SPS predictions with a limited impact of DPSs.

In the case of $J/\psi+W$ production, the observed yield of ATLAS~\cite{Aad:2014rua} also
ends up to be nearly one order of magnitude larger than the recent theory SPS evaluations 
from NRQCD (with CS and/or CO contributions)~\cite{Li:2010hc,Lansberg:2013wva} (see~\cite{Barger:1995vx,Kniehl:2002wd} for earlier studies).
 Just like the $J/\psi+Z$ case, the ATLAS raw-yield azimuthal distribution
 seems to exhibit a non-trivial structure hinting
at the presence of SPS events.

In this context, we have decided to have another look at prompt-$J/\psi+W$
 production at the LHC, in particular relying on an analysis of the SPS yield at
 NLO under the assumption of quark-hadron duality [the Colour-Evaporation Model
 (CEM)]~\cite{Fritzsch:1977ay,Halzen:1977rs} as in our earlier study~\cite{Lansberg:2016rcx}. As we
 demonstrated then, the CEM provides a conservative upper limit to the SPS
yield which we use to draw definite conclusions about the importance of the DPSs.

This paper is organised as follows.
In Section \ref{sec:JpsiW}, we briefly explain our NLO CEM SPS computation which,
as an upper theory limit for the SPS, allows us to claim for a clear indication of 
the contributions of DPSs at a 3 $\sigma$ level. In Section \ref{sec:DPSCEM}, we extract the DPS yield 
and determine $\sigma_{\rm eff}$ with its uncertainty range.
We then show that the azimuthal distribution of the prompt $J/\psi + W$ events collected
by ATLAS remains compatible with a yield highly dominated by DPSs.
Section \ref{sec:conclusion} is devoted to our conclusions and outlook.

\section{Our NLO evaluation of the SPS yield\label{sec:JpsiW}}

In this work, we thus focus on the production of a $W$ with a prompt $J/\psi$
 (which does not result from $b$-hadron decays) at the LHC in the kinematical
 region accessible by the ATLAS detector (see \ct{tab:phasespace}).

\begin{table}[htpb]
\begin{center} \footnotesize
\begin{tabular}{c|c}
\hline\hline
\multicolumn{2}{c}{$W$ boson selection\savefootnote{$M_T^W \equiv 2 P_T^{\mu^\pm} E_T \!\!\!\!\!\!\slash\ \ (1- \hbox{cos} (\phi^\mu -\phi^{\nu_\mu}))$
not to be confused with $m_T^2=m^2+P_T^2$.}}\\
\hline
\multicolumn{2}{c}{}\\
\multicolumn{2}{c}{$P_T^{\mu^\pm}$$>25$~{\rm GeV}, 
$E_T \!\!\!\!\!\!\slash ~> 20$ GeV,
$M_T (W) > 40$ GeV,
$|\eta_{\mu^\pm}|<2.4$}
\\
\multicolumn{2}{c}{}\\
\hline\hline
\multicolumn{2}{c}{$J/\psi$ selection}\\
ATLAS fiducial~\cite{Aad:2014rua} & ATLAS inclusive~\cite{Aad:2014rua} 
\\
\hline
$8.5<P_T^{J/\psi}<30\,{\rm GeV}$ & $8.5<P_T^{J/\psi}<30\,{\rm GeV}$ 
\\
$|y_{J/\psi}|<2.1$ & $|y_{J/\psi}|<2.1$ 
\\
$P_T^{\mathrm{leading~}\mu^\pm}>4.0$~{\rm GeV} &  
\\
$|\eta_{\mathrm{leading~}\mu^\pm}|<2.5$ & 
\\
either \ldelim({2}{5mm}$P_T^{\mathrm{sub-leading~}\mu^\pm}>2.5$~GeV \rdelim){2}{1mm}[] & 
\\
~~~~~~~~~~$1.3\leq |\eta_{\textrm{sub-leading~}\mu^\pm}|<2.5$  & 
\\
or ~~~~\ldelim({2}{5mm} $P_T^{\mathrm{sub-leading~}\mu^\pm}>3.5$~{\rm GeV} \rdelim){2}{1mm}[] &  
\\
~~~~~~~~~~$|\eta_{\textrm{sub-leading~}\mu^\pm}|<1.3$ & 
\\
\hline\hline
\end{tabular}
\caption{\label{tab:phasespace} Phase-space definition of the measured fiducial/inclusive 
production cross-section following the geometrical acceptance of the ATLAS 
detector.}
\end{center}
\end{table}

\flushfootnote

\begin{table*}[ht]
\begin{center}
\begin{tabular}{{c}*4 c} 
\hline\hline
Experiment \cite{Aad:2014rua} & LO CEM SPS & NLO CEM SPS & LO direct CSM SPS \cite{Lansberg:2013wva} & DPS \cite{Aad:2014rua} \\\hline
$12.6 \pm 3.2 \pm 0.9 ^{+4.1}_{-2.5}$ 
& $0.44^{+0.15}_{-0.10}$ 
& $0.78^{+0.21}_{-0.19}$ 
& $0.3 \pm 0.1$ & 72~{\rm mb}/$\sigma_{\rm eff}$ \\
\hline\hline
\end{tabular}
\caption{\label{tab:tot} 
Comparison of the experimental inclusive data for the normalised ratio
$\frac{d\sigma (pp \to W^\pm + J /\psi)}{dy_{J/\psi}}  \frac{{\rm BR}(J/\psi \to \mu^+ \mu^- )}{\sigma (pp\to W^\pm )}$ with our results of calculations in the CEM. We also show the normalised LO CS direct 
yield~\cite{Lansberg:2013wva} and the DPS yield ratios extracted from~\cite{Aad:2014rua}
as a function of $\sigma_{\rm eff}$. The unit is in $10^{-7}$. The theoretical 
uncertainty for the (N)LO SPS is from the renormalisation and factorisation scales. 
}
\end{center}
\end{table*}

As said above, the SPS contributions have been studied in the past within NRQCD, 
considering either CS~\cite{Lansberg:2013wva} or CO~\cite{Li:2010hc} channels. 
The latter CO evaluation at NLO accuracy in $\alpha_s$ is however perfectible 
as it relies on artificially low scales ignoring the emission of a $W$ boson
and with NRQCD Long Distance Matrix Elements (LDMEs) which 
are significantly larger than the NLO LDMEs compatible with the LHC yield and polarisation data 
(see \eg~\cite{Shao:2014yta}). 
That being said, since the direct CS yield  does not 
depend on free tunable parameters, it provides a strict {\it lower} limit on the 
SPS  $J/\psi +W$ production cross section.

On the contrary, the CEM most certainly sets an {\it upper} limit on the SPS 
yield. Indeed, the CEM tends to overshoot the experimental data for {\it single} 
quarkonium production at large $P_T^{J/\psi}$~\cite{Lansberg:2006dh,Brambilla:2010cs,Andronic:2015wma}\footnote{To 
cure this issue, different mechanisms~\cite{Edin:1997zb,Damet:2001gu,BrennerMariotto:2001sv,Ma:2016exq} 
were proposed but are not the object of a consensus. In our recent study~\cite{Lansberg:2016rcx}, we have shown that the NLO
corrections to the $P_T^{J/\psi}$ spectrum reduce the scale sensitivity but confirm the issue.}. This follows from the precocious 
dominance of the gluon-fragmentation topologies whose strength is therefore too 
high at large $P_T^{J/\psi}$. For $J/\psi+W$, the dominant CEM channel is 
quark-antiquark annihilation with the emission of a $W$ and an off-shell gluon 
fragmenting into a quarkonium. As such, the CEM likely results in an 
overestimation of the prompt $J/\psi+W$ cross section. It can thus be considered as 
a conservative upper limit of the theoretical prediction of the SPS yield. In 
other words, any set of NRQCD LDMEs which would be compatible with the yield 
predicted here would severely overshoot the very precise single-$J/\psi$ data. 
Let us also add that, due to the simplicity of the model, the CEM avoids 
the complexities encountered in NLO analyses of NRQCD, where 
a large theoretical uncertainty in any case remains. Overall, we will
consider that the difference between both (direct CSM \& CEM) represents our best 
evaluation of the theoretical uncertainty on the SPS cross section.

In the CEM, one considers the integral of the cross section for $Q \bar Q$ pair 
production with an invariant-mass between the quark mass threshold $2m_Q$ and 
that of open-heavy-flavour hadrons $2m_{H}$, where the hadronisation of heavy 
quarks into a quarkonium is likely.
To obtain the cross section to the hadron state ($\cal Q$), this integral is multiplied 
by a phenomenological factor for the probability of the quark states to hadronise 
into a given quarkonium state. In short, we would consider
\eqs{\sigma^{\rm (N)LO,\ \frac{direct}{prompt}}_{\cal Q}= \P^{\rm (N)LO,\frac{direct}{prompt}}_{\cal Q}\int_{2m_Q}^{2m_H} 
\frac{d\sigma_{Q\bar Q}^{\rm (N)LO}}{d m_{Q\bar Q}}d m_{Q\bar Q}.
\label{eq:sigma_CEM}}
$\P_{\cal Q}$ can be paralleled to the LDMEs in NRQCD. With the approach just described,
 the direct or prompt yields result from the same computation but with a different
 overall factor. In practice, we use $\P^{\rm LO, prompt}_{J/\psi}=0.014\pm 0.001$ 
and $\P^{\rm NLO, prompt}=0.009 \pm 0.0004$ which we have fit in~\cite{Lansberg:2016rcx} 
on the differential cross section for the production of single $J/\psi$ as measured by
ATLAS~\cite{Aad:2015duc} using $m_c=1.27$ GeV (see~\cite{Nelson:2012bc} for the mass choice). 
We refer to~\cite{Bedjidian:2004gd} for the first NLO CEM fits using the $P_T$-integrated
cross sections.

In view of \ce{eq:sigma_CEM}, computing $J/\psi+W$ production at NLO 
can be performed with modern tools of automated NLO frameworks, with some slight tunings.  
In practice, we have used {\small \sc MadGraph5\_aMC@NLO}~\cite{Alwall:2014hca}\footnote{We stress that, for the CEM, there is no
 need to use a specific automated tool like {\small \sc MadOnia}~\cite{Artoisenet:2007qm}
 and {\small\sc HELAC-Onia}~\cite{Shao:2012iz,Shao:2015vga} which are by the way
 currently not able to treat loop corrections.}.  

The hard scattering process to be considered is\footnote{Unlike the CSM case, 
channels such as $sg \to W^- J/\psi c$ negligibly contribute to the CEM yield and
can safely be neglected.}   $ij \to c \bar c + W^\pm + k$ 
with $i$, $j$ and $k$  standing for $g$, $q$ or $\bar q$. As what regards the
 parton distribution function (PDF), we have used the NLO 
NNPDF 2.3 PDF set~\cite{Ball:2012cx} with $\alpha_s(M_Z)=0.118$ provided by 
LHAPDF~\cite{Buckley:2014ana}. 

As already stated above, we have taken $m_c=1.27$~GeV, while choosing $m_c=1.5$~GeV 
would generate negligible changes to our results provided that the non-perturbative 
CEM parameter is chosen coherently. We note that the heavy-quark-mass dependence 
is de facto absorbed in the CEM parameter, hence the main theoretical uncertainties 
result from the renormalisation $\mu_R$ and factorisation $\mu_F$ scale variations 
which are believed to account for the unknown higher-order $\alpha_s$ corrections. 
In practice,  we have varied them independently within 
$\frac{1}{2}\mu_0\le \mu_R,\mu_F \le 2\mu_0$ where the central scale $\mu_0$ 
is the mass of $W$ boson, $M_W$. 

For the current analysis, ATLAS preferred to consider the ratio of the cross section
 for $W^\pm + J /\psi$ to that for $W^\pm$ in order to discard some efficiency and
 acceptance corrections related to the $W^\pm$ detection. For our theory evaluation,
 we use the NNLO $W$ boson production cross section in the ATLAS acceptance corrected
 by the branching ratio for the decay to muons $\sigma(pp\to W^\pm ) {\rm BR}(W \to \mu \nu) =5.08$ nb \cite{Aad:2011dm,Gavin:2010az,Gavin:2012sy}. In such a case, the $W^\pm + J /\psi$
 cross section in the numerator should also be multiplied by the branching
 ${\rm BR}(W \to \mu \nu)$ which cancels with that in the numerator. Overall
 one only needs to consider: 

\begin{equation}
\frac{d\sigma (pp \to W^\pm + J /\psi)}{dy_{J/\psi}} \frac{{\rm BR}(J/\psi \to \mu^+ \mu^- )}{ \sigma (pp\to W^\pm ) }
, 
\end{equation}
where  $y_{J/\psi}$ is the $J/\psi$ rapidity, ${\rm BR}(J/\psi \to \mu^+ \mu^- ) = 0.05961\pm 0.00033$ \cite{Olive:2016xmw}. We have also assumed 
${\rm BR}(W \to \mu \nu) = 0.1063$ and the $J/\psi$-rapidity range of 4.2 units to evaluate $d\sigma/dy_{J/\psi}$.

Our results (LO, NLO CEM) for the (inclusive) total cross section ratio 
are gathered in \ct{tab:tot}  along with the experimental results by ATLAS and 
the LO direct CS yield ratio computed in \cite{Lansberg:2013wva}. One notes that the 
NLO CEM ratio\footnote{On the way we note that the $K$ factor for our CEM evaluation 
is 1.77.} is --as expected-- nearly three times larger than the direct CSM ratio. In spite 
of the large experimental uncertainties, the SPS predictions strongly 
underestimate the measurements, by more than three standard deviations\footnote{In 
evaluating the discrepancy, we have not considered the spin uncertainties. Whereas, on general
grounds, the ATLAS evaluation is perfectly legitimate, we believe that, under the DPS
dominance hypothesis which we suggest, it probably 
lies outside the range allowed by the single $J/\psi$ polarisation measurement
of CMS~\cite{Chatrchyan:2013cla}, for instance.}. The same features 
can be observed in the distribution in $P_T^{J/\psi}$, shown in \cf{fig:ptspsdps} 
with the black (green) hatched histograms, resulting in a 3.1 standard-deviation 
discrepancy. As such we claim that these are clear indications of the existence of 
a DPS yield in this process which is 
supported by the analysis of the transverse momentum and azimuthal dependences
as we will show now.

\begin{figure}[hbt!]
\begin{center}
\includegraphics[width=\columnwidth]{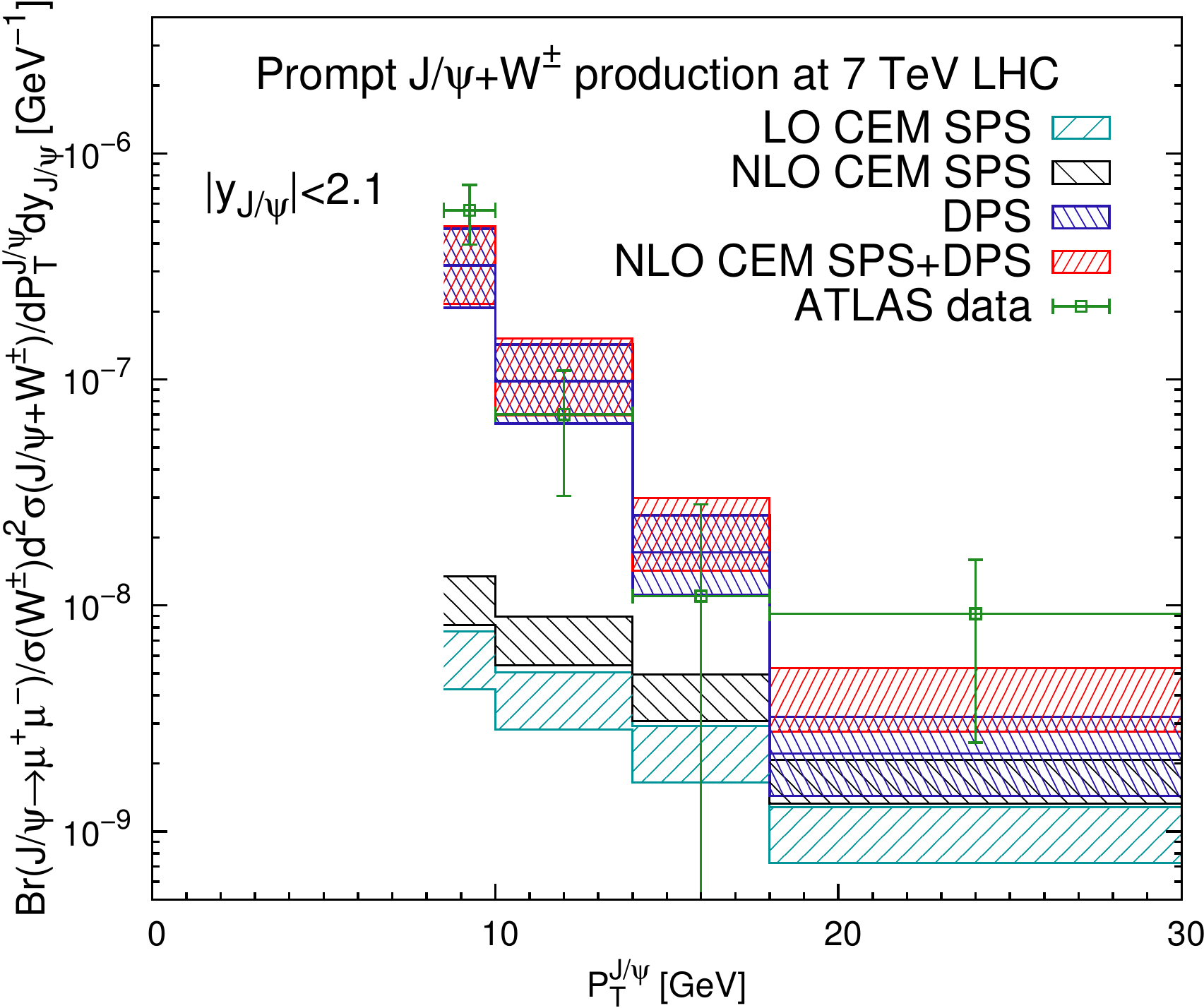}
\caption{ 
Comparison of $\frac{d\sigma (pp \to W^\pm + J /\psi)}{dy_{J/\psi}dP_T^{J/\psi}} \frac{{\rm BR}(J/\psi \to \mu^+ \mu^- )}{ \sigma (pp\to W^\pm ) }$ as a function of the transverse momentum of the prompt $J/\psi$. The division by $dy_{J/\psi}$ here means that we have divided the cross section by the rapidity interval of $|y_{J/\psi}|<2.1$. 
}
\label{fig:ptspsdps}
\end{center}
\end{figure}

\section{Our extraction of the DPS yield\label{sec:DPSCEM}}

\subsection{DPSs and the $P_T^{J/\psi}\!$-integrated cross section}

The DPS yield results from two uncorrelated scatterings within one proton-proton 
collision. As such, it is usually parametrised by the rudimentary pocket formula
which, for the process under study, reads
\begin{eqnarray}
d\sigma^{\rm DPS}(J/\psi+W)=\frac{d\sigma(J/\psi)d\sigma(W)}{\sigma_{\rm eff}}
,
\end{eqnarray}
where $\sigma_{\rm eff}$ is a supposedly universal parameter with the dimension
 of a surface. Recent experimental analyses are pointing at a visible impact
 of the DPSs in many reactions~\cite{Akesson:1986iv,Alitti:1991rd,Abe:1993rv,Abe:1997xk,Abazov:2009gc,Aad:2013bjm,Chatrchyan:2013xxa,Abazov:2014qba,Lansberg:2014swa,Aaboud:2016fzt}.

As a reference value, ATLAS used in their data-theory comparison~\cite{Aad:2014rua}
 $\sigma_{\rm eff} = 15 \pm 3 \hbox{(stat.)} ^{+5}_{-3} \hbox{(sys.)}$ mb (from
 $W+$ 2-jet data~\cite{Aad:2013bjm}) which results, 
{with $\sigma(J/\psi)$ and $\sigma(W)$ obtained from their
data}, in a normalised DPS cross section about 3 times smaller than
 their measurements (compare the first and last
 column of \ct{tab:tot}).

In the present analysis, we will instead extract the DPS yield by assuming 
that the difference between the ATLAS value and our SPS evaluation
is entirely due to DPSs. As such, the $\sigma_{\rm eff}$ associated to the present observable
 can also be constrained in a relatively precise manner\footnote{{Assuming the 
same single particle cross sections, $\sigma(J/\psi)$ and $\sigma(W)$, as
ATLAS}.}  and confronted to
other extractions. Let us recall that the SPS CEM evaluation should be regarded
as an upper limit of the SPS yield.

\begin{figure}[hbt!]
\begin{center}
\includegraphics[width=.8\columnwidth]{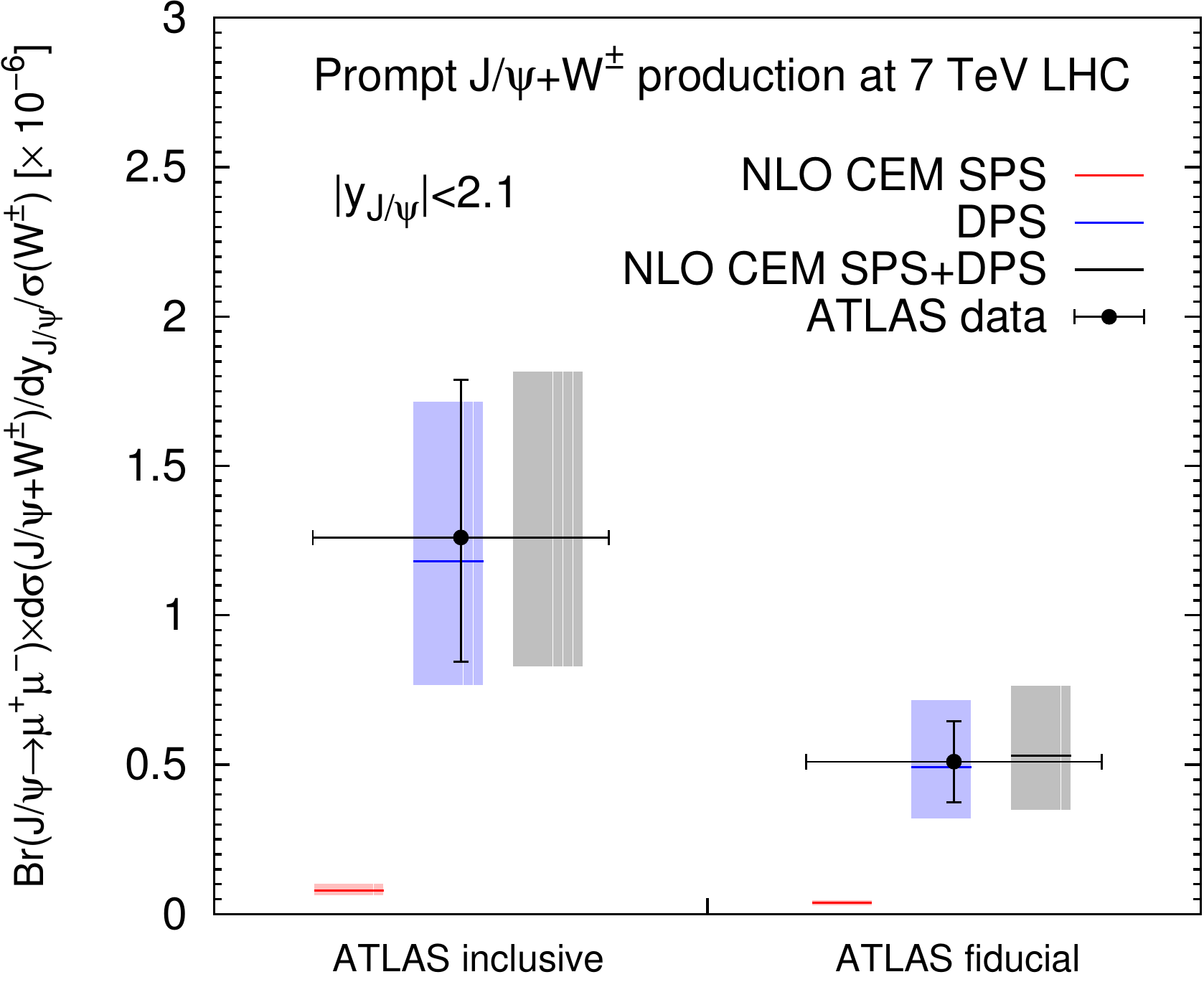}
\caption{Comparison of inclusive/fiducial $P_T^{J/\psi}\!$-integrated cross section ratios 
from the CEM NLO SPS, our fit DPS and their sum. The DPS uncertainties
reflect that of \ce{eq:sigma_eff}.}\label{fig:sigma_tot}
\end{center}
\end{figure}

\begin{table*}[hbt!]
\begin{center}
\begin{tabular}{cc|c} 
\hline\hline 
CEM, NLO & CSM, direct, LO \cite{Lansberg:2013wva}  & Combined\cr
\hline
$6.1^{+2.4}_{-1.3\, {\rm exp}} \, ^{+1.6}_{-1.6\, {\rm spin}} \,  ^{+0.1}_{-0.1\, {\rm theo}}$ 

& $5.8^{+2.2}_{-1.2\, {\rm exp}} \, ^{+1.5}_{-1.5\, {\rm spin}} \, ^{+0.05}_{-0.04\, {\rm theo}}$
& $6.1^{+3.3}_{-1.9\, {\rm exp}} \, ^{+0.1}_{-0.3\, {\rm theo}}$
\cr
\hline\hline
\end{tabular}
\caption{\label{tab:sigmaeff} 
Different extractions of $\sigma_{\rm eff}$ (in units of mb). The experimental uncertainties 
on our combination conservatively account for the ``spin" uncertainty (see the remark above).}
\end{center}
\end{table*}

\begin{figure*}[hbt!]
\begin{center}
\includegraphics[width=0.65\textwidth]{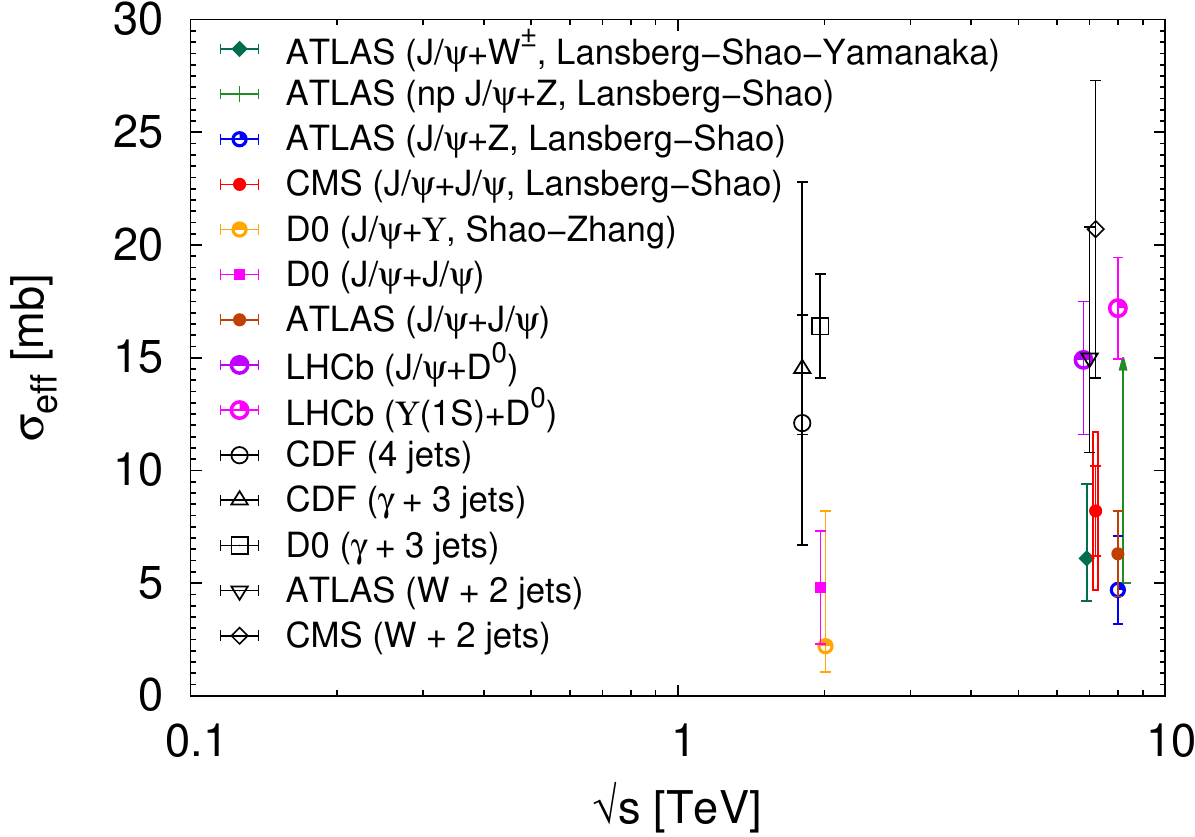}
\caption{
Comparison of our range for $\sigma_{\rm eff}$ 
($6.1^{+3.3}_{-1.9}$~mb)
 extracted from the $J/\psi+W$ data with 
other extractions~\protect\cite{Akesson:1986iv,Alitti:1991rd,Abe:1993rv,Abe:1997xk,Abazov:2009gc,Aad:2013bjm,Chatrchyan:2013xxa,Abazov:2014qba,Lansberg:2014swa,Aaboud:2016fzt,Aaij:2012dz,Aaij:2015wpa}.
\label{fig:sigma_eff}
}
\end{center}
\end{figure*}

Let us now describe how we evaluate $\sigma_{\rm eff}$ and its uncertainty.
Since the DPS yield is simply obtained by subtracting the SPS yield from 
the (inclusive) total cross section of ATLAS, the uncertainty on the DPS yield, and consequently
\footnote{{The uncertainties on the measured single-particle cross sections,  $\sigma(J/\psi)$ and $\sigma(W)$,
are negligible compared to that on $\sigma(J/\psi+W)$}.}
that of $\sigma_{\rm eff}$, will then depend on the (statistical and systematic) 
uncertainties of the data~\cite{Aad:2014rua} and on the range spanned by the SPS 
evaluations -- the NLO CEM and the direct LO CSM~\cite{Lansberg:2013wva}\footnote{We 
note that the NLO NRQCD evaluation~\cite{Li:2010hc} --despite the aforementioned drawbacks-- lies within this 
range.}. Since the SPS values are in both cases much smaller than the data, the theoretical 
uncertainties are in fact nearly irrelevant for the determination of 
$\sigma_{\rm eff}$. Our results are reported on~\ct{tab:sigmaeff} and on \cf{fig:sigma_tot}.

In particular, our combined result for $\sigma_{\rm eff}$ is then
\begin{equation}\label{eq:sigma_eff}
\sigma_{\rm eff}
=
(6.1^{+3.3}_{-1.9\, {\rm exp}} \, ^{+0.1}_{-0.3{\rm theo}} ) {\rm mb}
\end{equation}
which is nearly three times smaller than the ATLAS assumption (15 mb) with 
somewhat smaller uncertainties than found by other works. It is also 
consistent with our latest extraction from prompt $J/\psi +Z$ 
production~\cite{Lansberg:2016rcx} in \cf{fig:sigma_eff}, where we also compare our extraction with other measurements~\cite{Akesson:1986iv,Alitti:1991rd,Abe:1993rv,Abe:1997xk,Abazov:2009gc,Aad:2013bjm,Chatrchyan:2013xxa,Abazov:2014qba,Lansberg:2014swa,Aaboud:2016fzt,Aaij:2012dz,Aaij:2015wpa}.
Such a low value may hint at the non-universality of $\sigma_{\rm eff}$ in different processes, with a dependence on the flavour of the initial state, on the kinematics of the final state or on the energy of the proton-proton collision. For example, the 3 values of $\sigma_{\rm eff}$ from the ATLAS $J/\psi$-associated-production measurements are in general smaller than those from the LHCb measurements at forward rapidities. Such an observation seems to follow --at least qualitatively-- the lines of  the mean-field approximation~\cite{Blok:2016lmd}.

\subsection{Consistency with the transverse momentum and azimuthal distributions}

Our analysis shows that the DPSs are by far dominant and the SPSs
could even be omitted in the analysis of the $P_T^{J/\psi}\!$-integrated cross 
section. Since our procedure simply amounted to assume that any gap between the 
predicted SPS yield and the data was from DPS, it is important to check its 
consistency with differential cross sections. 

Looking back at \cf{fig:ptspsdps}, we see that the introduction of a somewhat 
larger DPS yield perfectly fills the gap where needed and does not create any 
surplus at large $P_T^{J/\psi}$ where the SPS was closest to the data. 
{We note that the plotted DPS $P_T^{J/\psi}$ spectrum follows from that of ATLAS~\cite{Aad:2014rua}
except for a different normalisation owing to the smaller value of $\sigma_{\rm eff}$
which we have just discussed. Like for the $P_T^{J/\psi}\!$-integrated DPS
cross section, ATLAS used the single-particle cross sections, here 
$d\sigma(J/\psi)/dP_T^{J/\psi}$, which they obtained from their own data.}

Indeed, the sum of the SPS and DPS yields, in red, gives a reasonable account of 
the ATLAS differential yield. This agreement should however not be 
over-interpreted in view of the large experimental uncertainties. This observation 
is rather a consistency check than a test. The good agreement at low 
$P_T^{J/\psi}$ simply follows from the fit of $\sigma_{\rm eff}$ to the total yield 
but resolves the 3-$\sigma$ discrepancy between ``theory" and ``experiment".

This also helps illustrate that the  low-$P_T^{J/\psi}$ yield is only from 
DPS contributions (as is the total yield) and that, at high $P_T^{J/\psi}$, DPS and 
SPS contributions could be of the same size. This is an important point to discuss 
the azimuthal dependence.

If initial-state-radiation effects are not too important, the SPS yield
(dominated by $2 \to 2$ scatterings) tends to peak at 
$\Delta \phi_{W\psi} \sim \pi$ --back-to-back scatterings-- whereas the DPS one 
is believed to be evenly spread in $\Delta \phi_{W\psi}$. \cf{fig:dphi} 
shows the event azimuthal distribution for prompt $J/\psi +W^\pm$ production. 
At first sight, it seems improbable that a yield largely dominated by DPS could be 
produced in agreement with such a distribution.

Just like for $J/\psi+Z$~\cite{Lansberg:2016rcx}, we note that the ATLAS acceptance for 
high-$P_T^{J/\psi}$ events is significantly higher than at low $P_T^{J/\psi}$. This, along with the very 
different SPS/DPS ratios as a function of $P_T^{J/\psi}$, can lead to a misinterpretation of such an
azimuthal dependence which is not corrected in acceptance.

In particular, the peak near $\Delta \phi_{W\psi} \sim \pi$, visible in the raw event 
distribution, could artificially be accentuated due to a large acceptance of high-$P_T^{J/\psi}$ events. 

Since we are not in the position of correcting the data, the simplest way to 
proceed is to mimic the folding of the theory event distribution with an approximated ATLAS acceptance. 
To do so, we use the $J/\psi$-acceptance dependence inferred in~\cite{Lansberg:2016rcx} where 
we assumed the background-over-signal ratio to be like $B/S \propto 1/P_T^{J/\psi}$. Reading out the
statistical uncertainties for 4 bins of \cite{Aad:2014rua}, we can derive the yield in each $P_T^{J/\psi}$ bin. 
In these, we can then compute the DPS-over-SPS ratio, derive the corresponding azimuthal dependence and finally 
combine them for the 4 bins using the yield in each. The resulting ``theory" distribution is shown on \cf{fig:dphi} and
agrees within uncertainties with the uncorrected ATLAS distribution.
On the way, we note that the number of SPS events following our computation at 
NLO CEM is $2\pm 1$ to be compared to $29^{+8}_{-7}$ events observed by ATLAS~\cite{Aad:2014rua}. 
We expected these $2\pm 1$ SPS events to lie at $\Delta \phi_{W\psi} \sim \pi$. 
Conversely, we expect an updated ATLAS $\Delta \phi_{W\psi}$ analysis in 2 $P_T^{J/\psi}$ bins 
to show a flat behaviour in the ``low" $P_T^{J/\psi}$ bins and a slightly peaked one
for the ``high" $P_T^{J/\psi}$ bins. It however seems that more statistics is needed to draw
final conclusions.

\begin{figure}[hbt!]
\begin{center}
\subfloat[NLO]{\includegraphics[width=\columnwidth]{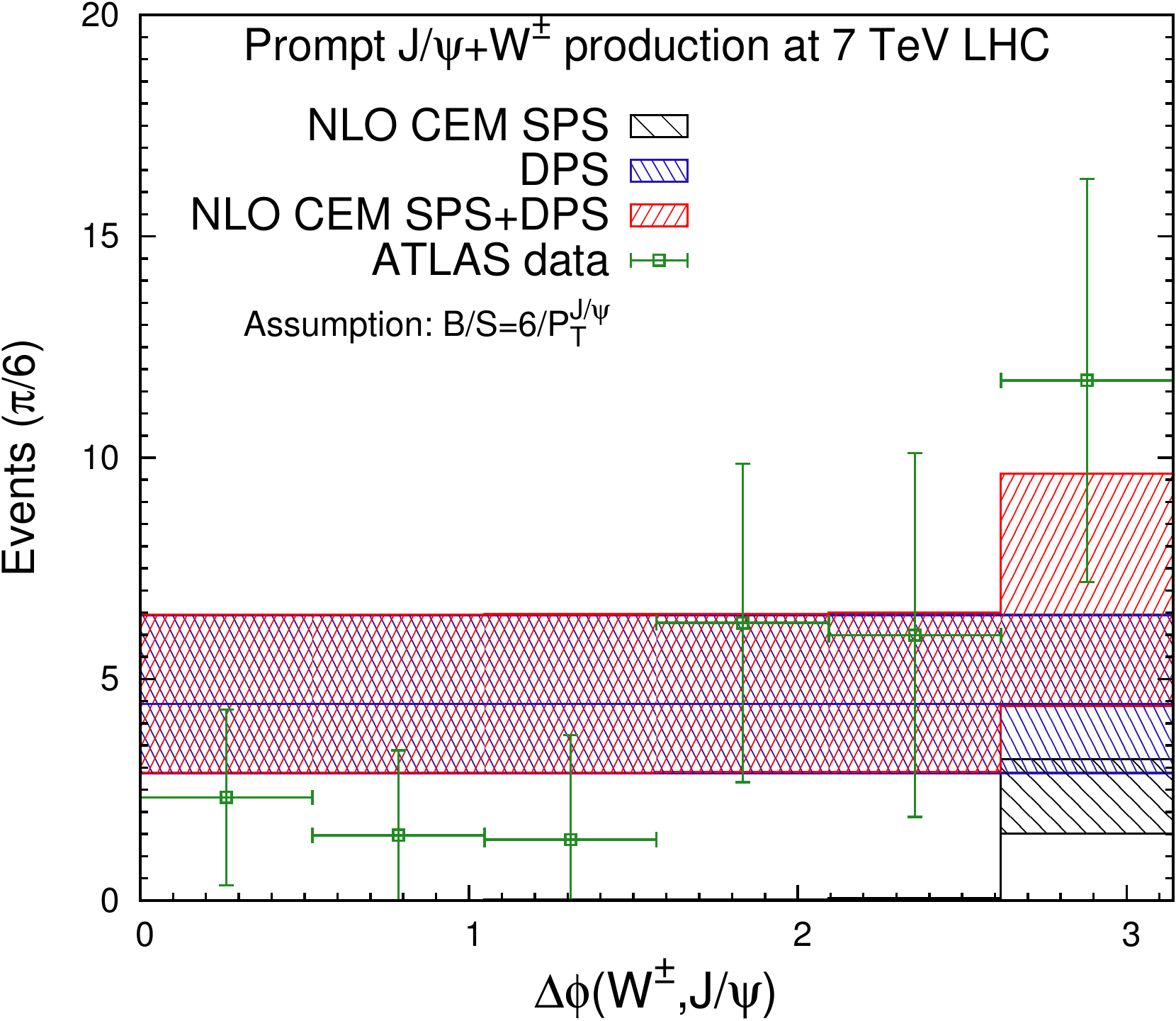}}
\caption{
Comparison between the (uncorrected) ATLAS azimuthal event distribution and our
 NLO theoretical results for $J/\psi+W$ in the CEM (SPS + DPS) effectively
 folded with an assumed ATLAS acceptance.
\label{fig:dphi}}
\end{center}
\end{figure}

\section{Conclusions\label{sec:conclusion}}

We have re-analysed the associated production of a prompt $J/\psi$ with a $W$
 boson at the LHC in view of the ATLAS data~\cite{Aad:2014rua}. To do so, we
 have performed the first NLO calculation of the SPS cross sections in the CEM
 which we consider to be a conservative upper limit of the SPS yield. This has
 allowed us to claim a 3-$\sigma$ deviation between SPS theory and the ATLAS
 data, which we interpret as a strong indication for DPSs. Our conclusion does not
 depend on the quarkonium-production model used to compute the SPS yields.

We have then determined  a finite range for $\sigma_{\rm eff}$ around 6 mb,
 consistent with extractions from other experiments and other quarkonium-related
 processes. In particular, we emphasise that our result is consistent with
 $\sigma_{\rm eff}$ extracted from our analysis of prompt $J/\psi +Z$ production
 as measured by ATLAS.

Whereas the azimuthal dependence could be very useful in separating the SPSs
 from the DPSs, one has to be very cautious with uncorrected
 raw event distribution. A further analysis with higher statistics allowing
one to study the rapidity-separation spectrum and a fully
 corrected azimuthal distribution will further constrain the range for
 $\sigma_{\rm eff}$ since the theoretical uncertainties on the SPS happen to be
 quasi irrelevant at low $P_T^{J/\psi}$.


\section*{Acknowledgements} 
We thank V. Kartvelishvili and D. Price for useful discussions. 
The work of J.P.L. is supported in part by the French IN2P3--CNRS via the LIA FCPPL (Quarkonium4AFTER) and the project TMD@NLO. 
H.S.S. is supported by the ILP Labex (ANR-11-IDEX-0004-02, ANR-10-LABX-63).
N.Y. is supported by a 	JSPS Postdoctoral Fellowships for Research Abroad and by then Riken iTHES Project.

\bibliographystyle{utphys}

\bibliography{onium+W-CEM-010518}

\appendix

\end{document}